\documentclass[12pt, a4paper]{article}  
\usepackage{amsmath}
\usepackage{typearea}
\typearea{12}
\input{epsf.tex}
\allowdisplaybreaks
\begin{document}
\newcommand{\bi}[1]{\bibitem{#1}}
\def\be#1\ee{\begin{equation}#1\end{equation}}
\def\bea{\begin{eqnarray}}
\def\eea{\end{eqnarray}}
\def\6{\partial} \def\a{\alpha} \def\b{\beta}
\def\g{\gamma} \def\d{\delta} \def\ve{\varepsilon} \def\e{\epsilon}
\def\z{\zeta} \def\h{\eta} \def\th{\theta}
\def\vt{\vartheta} \def\k{\kappa} \def\l{\lambda}
\def\m{\mu} \def\n{\nu} \def\x{\xi} \def\p{\pi}
\def\r{\rho} \def\s{\sigma} \def\t{\tau}
\def\Ph{\phi} \def\ph{\varphi} \def\ps{\psi}
\def\o{\omega} \def\G{\Gamma} \def\D{\Delta}
\def\Th{\Theta} 
\def\Lam{\Lambda} 
\def\S{\Sigma}
\def\PH{\Phi} \def\Ps{\Psi} \def\O{\Omega}
\def\sm{\small} \def\la{\large} \def\La{\Large}
\def\LA{\LARGE} \def\hu{\huge} \def\Hu{\Huge}
\def\ti{\tilde} \def\wti{\widetilde}
\def\non{\nonumber\\}
\def\rG{{\buildrel {\{\}} \over \G}}
\def\rR{{\buildrel {\{\}} \over R}}
\def\semidirect{\;{\rlap{$\subset$}\times}\;}
\def\xt{{\tilde x}}
\def\FF{{\cal F}}
\def\GG{{\cal G}}

\title{On Measuring Gravitomagnetism via Spaceborne Clocks:
       A Gravitomagnetic Clock Effect}
\author{B.Mashhoon${}^{1,2}$, F.Gronwald${}^{2}$, F.W.Hehl${}^2$, 
and D.S.Theiss${}^2$ 
\\ \\ \\
${}^1$Department of Physics and Astronomy\\
University of Missouri, Columbia, Missouri 65211\\
USA\\
\\${}^2$Institute for Theoretical Physics
 \\University of Cologne, 50923 K\"oln\\ GERMANY}
\date{}
\maketitle

{\footnotesize
\abstract{The difference in the 
proper azimuthal periods of revolution of two standard clocks in direct and 
retrograde orbits about a central rotating mass is proportional to $J/Mc^2$,
where $J$ and $M$ are, respectively, the proper angular momentum and mass
of the source. In connection with this gravitomagnetic clock effect, we explore
the possibility of using spaceborne standard clocks for detecting the 
gravitomagnetic field of the Earth. It is shown that this approach to the 
measurement of the gravitomagnetic field is, in a certain sense, theoretically
equivalent to the Gravity Probe$\,$-B concept.}}

\vfill
\pagebreak

\section{Introduction}

Currently there is considerable interest in the development of highly stable 
clocks for space applications \cite{mash96}. This circumstance provides the 
impetus to investigate further a certain remarkable gravitomagnetic clock 
effect \cite{cohe93} within the framework of general relativity.

The Newtonian theory of gravitation may be thought of as the nonrelativistic
theory of the gravitoelectric field ${\bf E}_g$, 
while general relativity involves -- among other things -- the 
gravitomagnetic field ${\bf B}_g$ as well. These notions
are ultimately based on the close formal analogy between Newton's law of 
universal gravitation and Coulomb's law of electricity. The gravitomagnetic 
clock effect involves a certain characteristic temporal structure around
rotating bodies. The elucidation of the various aspects of this effect 
is the main subject of this paper; in particular, we consider the theoretical 
problem of measuring the non-Newtonian gravitomagnetic field by means of 
ultra-stable clocks in space.

The standard tests of the Einstein theory of gravitation can be accounted for
by post-Newtonian gravitoelectric corrections: gravitational redshift,
perihelion precession of Mercury, bending of light in the field of the 
Sun, and Shapiro's radar time delay. The investigation of gravitomagnetic 
effects in general relativity began with the work of de Sitter, Thirring, and 
Lense and Thirring \cite{mash84}. However, the idea of a gravitomagnetic field
generated by mass current dates back to the last century when developments 
in electromagnetism suggested a generalization of Newton's theory of 
gravitation along the lines of electrodynamics \cite{nort,
whit}. Holzm\"uller \cite{holz70} and Tisserand \cite{tiss72} tried to
explain the perihelion excess of Mercury by taking into account solar
gravitomagnetism. The analogy with electrodynamics has been the
subject of many studies \cite{wiec20, hund84, scia53}; in fact, any
theory that brings together Newtonian gravitation and Lorentz
invariance must contain gravitomagnetism in some form \cite{natu}. At
present, general relativity is consistent with all observational data and
within its framework the theoretical development of gravitoelectromagnetism
has reached a certain level of maturity \cite{mash97}. On the experimental
side, moreover, the main mission of NASA's Gravity Probe$\,$-B will be the
direct measurement of the gravitomagnetic field of the Earth via
superconducting gyroscopes carried by a drag-free satellite in a
polar orbit about the Earth \cite{ever86}.

In this paper, we choose an astronomical body such as the Earth for the
sake of concreteness and concentrate our attention on the theoretical
possibilities for measuring its gravitomagnetism by means of
spaceborne clocks. To this end, the exterior gravitational field of
the Earth may be considered in the linear approximation with a metric of the
form
\be
-ds^2\,=\, -c^2\Bigl(1-{\frac{2GM}{c^2\rho}}\Bigr)dt^2 + 
\Bigl(1+{\frac{2GM}{c^2\rho}}\Bigr)
\d_{ij}dx^i dx^j-{\frac{4G\, dt}{c^2 \rho^3}}
\e_{ijk}J^i x^j\, dx^k \,,
\ee
where $M$ and $J$ are the mass and angular momentum of the body,
respectively. Here $\rho$ is the isotropic radial coordinate in an
underlying Cartesian coordinate system ${\boldsymbol \rho}=(x^i)=(x,y,z)$ 
and we 
choose an orientation such that ${\bf  J}=J{\bf \hat{z}}$. The
Lense-Thirring part of this metric has the form $-4c^{-1}({\bf A}_g\cdot
d{\bf x})dt$, where
\be
{\bf A}_g\,=\,\frac{G}{c} \frac{{\bf J}\times{\boldsymbol \rho}}{\rho^3}
\ee
is the gravitomagnetic vector potential. A free ideal gyroscope held
at rest at a fixed position ${\boldsymbol \rho}$ in space would precess in the
gravitomagnetic field at a rate given by
\be
{\boldsymbol \Omega}_P\,=\, 
{\frac{G}{c^2\rho^5}}\Bigl[3({\bf J}{\boldsymbol{\cdot\rho}})
{\boldsymbol \rho} - \rho^2{\bf J}\Bigr]  \label{precess}
\ee
in this approximation. The gravitomagnetic field is then defined by
${\bf B}_g=c\,{\boldsymbol \Omega}_P={\bf \nabla}\times{\bf A}_g$, while the
gravitoelectric field is ${\bf E}_g=-{\bf \nabla}\Phi_N$, where 
$\Phi_N=GM/\rho$ is the Newtonian potential, as expected.

In the following sections 2 and 4, the Schwarzschild radial coordinate
$r$ is employed,
\be
r=\rho\Bigl(1+\frac{1}{2}{\frac{GM}{c^2\rho}}\Bigr)^2\,;
\ee
moreover, we usually set $G=1$ and $c=1$ for the sake of simplicity
-- except where indicated otherwise. The gravitomagnetic 
clock effect under consideration 
in this paper is described in terms of azimuthal closure for simple
geodesic orbits in sections 2 -- 4. This gravitomagnetic effect may
be thought of in terms of a certain limiting form of the gravitational
Aharonov--Bohm effect; indeed, this connection leads to a discussion 
of holonomy in section 5. A preliminary examination of the experimental 
possibilities is provided in section 6. The measurement of the gravitomagnetic
clock effect appears to be beyond present experimental capabilities
by almost an order of magnitude. 

\section{A Gravitomagnetic Clock Effect} \label{sec2}
A standard clock by definition measures proper time along its
worldline. Imagine such clocks in the stationary spacetime outside
a charged rotating source; in fact, we take this region to be the 
exterior Kerr--Newman spacetime for the sake of simplicity.

Consider equatorial circular geodesic orbits in this spacetime. We are 
interested in timelike orbits that are stable against radial perturbations.
This is possible for circular orbits only beyond a certain radius.
Let $t_+$ $(t_-)$ be the period of revolution in the same (opposite) sense as 
the rotation of the source for the stable circular geodesic path with fixed 
radial coordinate $r$ in standard Schwarzschild-like coordinates.  It can be
shown that 
\be
t_{\pm}\,=\,T_0 \pm 2\pi\frac{a}{c}  \,, \label{exact2}
\ee
where $T_0=2\pi/{\omega_0}$,~ $\omega_0$ is the modified ``Keplerian'' 
frequency given by
\be
\omega_0\,=\,\Bigl({\frac{GM}{r^3}}-{\frac{GQ^2}{c^2r^4}}
\Bigr)^{\frac{1}{2}}\,,
\ee
$Q$~denotes the charge, and $a=J/Mc$ is the Kerr parameter \cite{equa}. 
The coordinate time $t$ is the proper time of static asymptotically 
inertial observers that are infinitely far from the source. It follows
from equation (\ref{exact2}) that
\be
t_+ - t_- \,=\, 4\pi\frac{J}{Mc^2}\,.
\label{exact3}
\ee
The operational significance of this interesting result is doubtful,
since light signals are required, for instance, to carry information about
the orbit to the distant clocks. To ameliorate this situation, let us
consider instead the {\it proper} periods, $\tau_\pm$, of such circular paths 
measured by free orbiting standard clocks. It is possible to show that
\cite{equa}
\be
\tau_{\pm}\,=\,T_0\Bigl[1-{\frac{3GM}{c^2 r}}+2{\frac{GQ^2}{c^4 r^2}}\pm
2\frac{a}{c}\omega_o\Bigr]^{\frac{1}{2}}\,, \label{exact}
\ee
so that
\be
{\frac{1}{2T_0}}\Bigl(\tau_+^2-\tau_-^2\Bigr)\,=\,4\pi {\frac{J}{Mc^2}}\,;
\label{differ}
\ee
hence, an experimental determination of the left side of equation 
(\ref{differ}) would lead to the measurement of $J/M$ for the
source. Specifically, let us write equation (\ref{differ}) as
\be
\tau_+ - \tau_- \,=\, 4\pi{\frac{J}{Mc^2}} U(r)\,,
\ee
where $U^{-1}\equiv \frac{1}{2}(\tau_+ + \tau_-)/T_0$ and $U(r)$ approaches 
unity as $r\rightarrow\infty$. Under physically realistic conditions,
$U$ is a monotonically decreasing function of $r$; in fact, for 
\be
\Phi\equiv{\frac{GM}{c^2 r}} \ll 1  \, ,
\ee
$U(r)$ has an expansion of the form
\be
U(r) = 1+\frac{3}{2}\Phi + \Bigl(\frac{27}{8} - {\frac{Q^2}{GM^2}}
\Bigr)\Phi^2+
\Bigl(\frac{135}{16}+ \frac{1}{2}{\frac{c^2J^2}{G^2 M^4}} - {\frac{9}{2}}
{\frac{Q^2}{GM^2}}\Bigr)\Phi^3 + \cdots \quad .
\ee

The exact result (\ref{exact}) is valid for the Kerr--Newman geometry.
We are interested, however, in the exterior field of a rotating
astronomical body; therefore, $r \gg GM/c^2$ and hence
\be
\tau_+ - \tau_- \,\simeq\, 4\pi\frac{J}{Mc^2}\,. \label{general}
\ee
This is a remarkable relation that could, in principle, be used to
measure $J/M$ directly for an astronomical body; in fact, for the
Earth $\tau_+ - \tau_- \simeq 2\times 10^{-7}{\rm sec}$, while for
the Sun $\tau_+ - \tau_- \simeq 10^{-5}{\rm sec}$. Equations (\ref{exact3})
and (\ref{general}) indicate a general feature of time in the field of
a rotating source; this paper is devoted to an account of this gravitomagnetic
clock effect and its possible observational significance. 

The general result that it takes longer for a free test particle to go
around a rotating mass in its equatorial plane in the prograde direction than
in the retrograde direction is a remarkable fact that is in conflict with
the notion that a rotating mass drags space around with it; in fact,
such a ``Machian'' concept is in conflict with general relativity and
must be abandoned \cite{mash84, rind97}. Moreover, the ``dragging of local 
inertial frames'' has often been used as a metaphor for the gravitomagnetic 
precession of ideal test gyroscopes; however, it must be remarked that 
even this figurative usage has the erroneous connotation just mentioned.
There are two other interrelated aspects of the 
general relation (\ref{general}) that
are quite interesting and require further discussion. First, it is
clear from equation (\ref{general}) that $\tau_+ - \tau_-$ is nearly
independent of Newton's gravitational constant $G$. 
Intuitively, this essentially comes about since {\it in this approximation} 
a ``small'' quantity is integrated over
a ``long'' interval. Thus the result could be a ``large'' effect,
since it is independent of the extremely weak gravitational coupling
constant in this limit. It is useful to recall here a similar
situation involving the total gravitational radiation energy emitted
when a test particle of mass $m$ that is at rest at infinity falls
radially into a Schwarzschild black hole of mass $M \gg m$; the net
result, $\simeq 10^{-2}(m/M) mc^2$, is also independent of
$G$. Another analogous circumstance involves the net amplitude 
of {\it relativistic nutation} that is independent of $c$; that is,
the spin vector of a test particle in an orbit of inclination $\a$ 
about a rotating mass undergoes -- in addition to the normal precessional
motions -- a certain nutational motion of long (Fokker) period
$T_F\simeq 2 T_0/3\Phi$ with an amplitude $J\sin\a/Mr^2\omega_0$ that
is independent of $c$ \cite{mash85}. This is a consequence of the
post-Schwarzschild approximation scheme that is discussed in section
4.

The other significant aspect of the result (\ref{general}) is that it
is essentially independent of $r$ for $r \gg 2GM/c^2$. This
gravitomagnetic clock effect is thus reminiscent of the topological
Aharonov-Bohm effect. The connection between these effects can be
further clarified as follows: The Aharonov-Bohm effect is simply
related to the Sagnac effect via the Larmor theorem \cite{saku};
in a similar way, the gravitational Larmor theorem provides a
connection between the gravitomagnetic effect (\ref{general}) and
an analog of the Sagnac effect. This is discussed in the next
section.

\section{Analogy with the Sagnac Effect}

A century ago, Larmor's theorem provided a local connection between magnetism
and rotation. The theorem applies to circumstances involving slowly moving
charged particles and slowly varying fields whose strengths are considered
only to first order. It turns out (cf.\ Appendix A) that the theorem can be
extended to the gravitational case such that the gravitoelectric charge 
of a test mass $m$ would be $q_E = -m$ and its gravitomagnetic charge would
be $q_B = -2m$. This is the content of the gravitational Larmor theorem
\cite{mash93}; the negative signs of the gravitational charges account for
the attraction of gravity and $q_B/q_E =2$ since gravity is a spin-2 field.

It follows from Larmor's relation
\be
{\boldsymbol \Omega}_L = {\frac{q}{2mc}} {\bf B}
\ee
that in the gravitomagnetic case with $q_B = -2m$, we have 
\be
{\boldsymbol \Omega}_L = -{\boldsymbol \Omega}_P\,,
\ee
where ${\boldsymbol \Omega}_P = {\bf B}_g/c$ is 
the local gravitomagnetic precession 
frequency of a gyroscope at rest. It follows from equation (\ref{precess})
that in the equatorial plane $({\bf J}{\boldsymbol{\cdot \rho} =0})$ 
one should obtain 
the same physical result at an orbit of radius $\rho$ in the absence of 
gravity but in a frame rotating with uniform frequency 
${\boldsymbol \Omega}_L= G{\bf J}/c^2\rho^3$. 
Imagine, therefore, two clocks moving
in opposite directions with speed $v$ on the circular orbit of radius 
$\rho$ in the absence of gravity. According to an observer at rest in the 
frame rotating with frequency ${\bf \Omega}_L$, the periods of circular 
motion for the two clocks are
\be
t_{\pm}\,=\, {\frac{2\pi\rho}{v\mp \rho \Omega_L}}
\ee
by the nonrelativistic law for the addition of velocities. (The same result
can be obtained with respect to static observers in the underlying 
inertial frame.) It follows that
\be
t_+ - t_- \,=\, {\frac{4(\pi\rho^2)\Omega_L}{v^2 - \rho^2\Omega_L^2}}\,,
\label{differ2}
\ee
which for $v \gg \rho\Omega_L$ and $v\rightarrow c$ reduces to the Sagnac 
effect for light -- though the derivation presented here breaks down,
of course, as it is only valid in the nonrelativistic approximation.
That is, the Sagnac phase shift may be thought of -- in the eikonal 
approximation -- as the product of frequency of light and the total time
difference for light to traverse the circular area in opposite directions. 
As is well known, the Sagnac effect is proportional to the enclosed
area $(\pi \rho^2)\,$; in fact, the magnitude of the Sagnac phase shift is
given in our case by $4\omega^*(\pi\rho^2)\Omega_L/c^2$, where $\omega^*$ is
the frequency of the electromagnetic radiation.

Let us now compute equation (\ref{differ2}) for the gravitomagnetic case
$v=\rho\,\omega_0$, where $\omega_0$ is the Keplerian frequency,
$\omega_0^2=GM/\rho^3$, and 
\be
{\frac{\Omega_L^2}{\omega_0^2}}=\frac{J}{Mc^2}\Omega_L \ll 1 
\ee
in all physically realistic situations. Hence we find that
\be
\tau_+ - \tau_- \,\simeq \,4\pi{\frac{\Omega_L}{\omega_0^2}} \,=\, 4\pi 
\frac{J}{Mc^2}\,,
\ee
as expected from the gravitational Larmor theorem.

The simplicity of the above argument is due to the fact that at a given $\rho$
in the equatorial plane ${\bf \Omega}_L$ is uniform; however, for an orbit
at an inclination $\a\neq 0$ this would no longer be the case. A more 
significant problem is that an orbit in the field of a rotating mass is not
in general spatially closed. In fact, the equatorial circular geodesic
orbits discussed thus far are exceptional in this respect. It is therefore 
necessary to formulate a generalization of relation (\ref{general}) for
an arbitrary orbit; this problem is treated in the next section.  

\section{Standard Clocks in Space}
A spaceborne clock would in general follow a complicated orbit. Thus far, we 
have considered only circular geodesic orbits in the equatorial plane 
of a rotating mass; however, it is possible to generalize the
gravitomagnetic clock effect to the case of an arbitrary orbit.
To clarify the situation, we first consider a ``spherical'' orbit that
has a small inclination $\a \ll 1$ with respect to the equatorial plane.
Spherical orbits in Kerr spacetime have been described by Wilkins \cite{wilk}.
Such a geodesic orbit is no longer spatially closed in general; therefore, it
is necessary to define the relevant periods $\tau_\pm$ in terms of 
{\it azimuthal closure}.

Most elementary astronomical systems, i.e. stars and planets, are nearly 
spherical bodies. The exterior gravitational field of a spherically symmetric 
mass distribution is uniquely described by the Schwarzschild spacetime; 
therefore, it is natural to express the exterior field of such an astronomical
system in terms of perturbations of the Schwarzschild field. The 
post-Schwarzschild approximation scheme is a useful method in relativistic 
celestial mechanics \cite{mash85, thei84, mash82}. It is possible, in
principle, to include the quadrupole and higher mass moments of the source
in our treatment; however, the analysis of the gravitomagnetic clock
effect for a general orbit would then become much more complicated 
and is beyond the scope of this paper. The unique first-order 
angular momentum perturbation is simply given by the Kerr field linearized in 
the angular momentum parameter $a$. Thus we express the exterior field of a 
rotating mass by
\be
-ds^2 \, = \, -c^2(1-2\Phi)dt^2 + (1-2\Phi)^{-1}dr^2 + r^2(d\theta^2+
           \sin^2\theta\, d\varphi^2)-
           4cr\Psi\sin^2\theta\, dt\,d\varphi\,,
\ee
where $\Psi=GJ/c^3r^2$ is a gravitomagnetic potential. For the exterior field 
of the Earth, $\Phi(r) < 7\times 10^{-10}$ and $\Psi(r) < 4\times 10^{-16}$.

Let us now consider a ``spherical'' orbit of small inclination $\a \ll 1$
given by
\bea
t&=&\Gamma_0\Bigl(1-3\frac{a\omega}{c}\Gamma_0\Phi_0\Bigr)\tau \,,\label{ini1}
 \\
r&=& r_0\,,\\
\theta&=& {\frac{\pi}{2}}-\a\sin\h + 3\a{\frac{a\omega}{c}}\Gamma_0(1-2\Phi_0)
\omega\tau\cos\h\,, \\
\varphi&=& \Bigl(1-\frac{a\omega}{c}\Gamma_0\Bigr)\omega\tau + \varphi_0\,, 
\label{ini4}
\eea
where $\Phi_0 = \Phi(r_0)$, $\omega$~is the proper frequency of the 
unperturbed orbit, $\omega = \omega_0\Gamma_0$\,,
\be
\Gamma_0 = (1-3\Phi_0)^{-\frac{1}{2}}
\ee
and $\h$ is -- for the unperturbed orbit -- the phase angle in the orbital 
plane measured from the line of nodes
\be
\h\,=\,\omega\tau + \h_0\,.
\label{nodes}
\ee
The constants $\varphi_0$ and $\h_0$ are given in terms of initial conditions
at $\tau=0$ as in Figure 1. The linear perturbation analysis is valid for
$\omega_0\tau \ll (a\omega_0/c)^{-1}$. It is possible to show that the average
behavior of the orbit is described by the Lense--Thirring effect with a 
precession frequency $\Omega_{LT}\simeq 2c\Psi(r_0)/r_0$.

\begin{figure}[htb]
\centerline{\epsfysize=80mm\epsffile{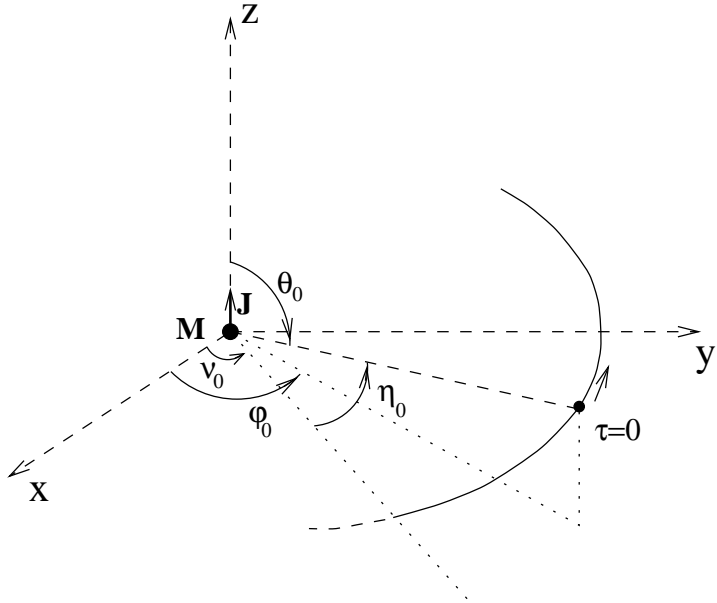}}
\caption{Plot of the direct ``spherical'' orbit of radius $r_0$. The azimuth of
the line of nodes $\nu_0$ is given by $\nu_0=\varphi_o-\arctan(\cos\alpha
\tan\eta_0)$. The polar angle at $\tau = 0$ is given by $\cos\theta_0 =
\sin\alpha\sin\eta_0$. Here $\alpha$ is the inclination of the 
unperturbed (circular) orbit with respect to the equatorial plane 
of the rotating mass.}
\end{figure} 
\bigskip
\bigskip

It is clear from equations (\ref{ini1}) -- (\ref{ini4}) that the orbit is
no longer spatially closed once $\a \neq 0$. Let us define $\tau_+$
to be the period of motion from $\varphi_0$ to $\varphi_0 + 2\pi$; hence,
we obtain from equation (\ref{ini4}) 
\be
\Bigl(1-\frac{a\omega}{c}\Gamma_0\Bigr)\omega \tau_+ \,=\, 2\pi\,.
\ee
It follows from this relation  that $(T=2\pi/\omega)$
\be
\tau_+\,=\, T+2\pi\Gamma_0 \frac{a}{c}\,;
\ee
moreover, for a retrograde orbit, $\,a\rightarrow -a\,$ so that
\be
\tau_- \,=\, T-2\pi\Gamma_0\frac{a}{c}\,.
\ee
Hence, we recover $\tau_+ - \tau_- \simeq 4\pi a/c$ for
$\Phi_0 \ll 1$. Thus the rotation of the source breaks the degeneracy in the
proper period $T$ of the orbit by an amount that is of the order of
$J/Mc^2$. This is illustrated in Figure 2.

\begin{figure}[htb]
\centerline{\epsfysize=80mm\epsffile{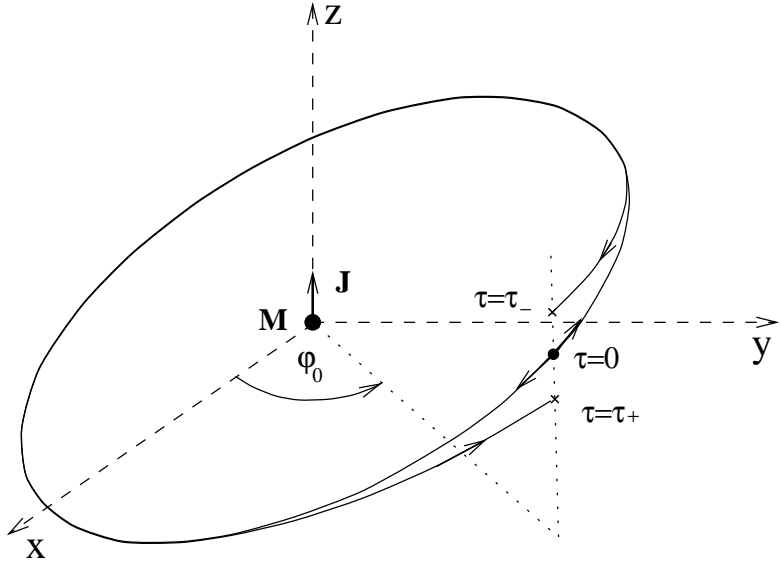}}
\caption{Schematic plot of clock orbits about a rotating mass. The
azimuthal periods are illustrated for $\cos\eta_0 > 0$; in fact, the 
deviation in the polar angle after one orbit is 
$4\pi\alpha a\cos\eta_0/c\Gamma_0$. }
\end{figure} 
\bigskip
\bigskip

Let us next consider a general ``spherical'' orbit of arbitrary inclination
$\a$; in fact, the orbital equations have been derived in previous work 
\cite{thei84}. We have
\bea
t(\tau)&=& \Gamma_0\Bigl(1-3\frac{a\omega}{c}\Gamma_0\Phi_0\cos\a\Bigr)\tau\,,
\label{2ini1} \\
r(\tau)&=& r_0\,, \\
\theta(\tau)&=& \arccos(\sin\a\sin\h)+{\frac{3}{2}}\frac{a\omega}{c}
\Gamma_0(1-2\Phi_0)\sin(2\a) {\frac{\cos\h}{\sigma(\h)}}\omega\tau\,, \\
\varphi(\tau)&=& \arctan(\cos\a\tan\h)+{\frac{a\omega}{c}}
\Bigl[2\Gamma_0^{-1}-3\Gamma_0(1-2\Phi_0){\frac{\cos^2\a}{\sigma^2(\h)}}\Bigr]
\omega\tau + \nu_0\,,  \label{2ini4}
\eea
where $\h$ is defined as in equation (\ref{nodes}), $\sigma(\h)$ is given by 
\be
\sigma(\h)\,=\, (1-\sin^2\a\sin^2\h)^{\frac{1}{2}} \,,
\ee
and $\nu_0$ is the longitude of the node as in Figure 1. For $\a\ll 1$,
equations (\ref{2ini1}) -- (\ref{2ini4}) reduce to the previous orbital
equations (\ref{ini1}) -- (\ref{ini4}); moreover, the Lense--Thirring 
precession as well as other properties of the general orbit have been 
described in detail before \cite{thei84}. For $\a = \pi/2$, the main orbital
motion is polar and the azimuthal motion indicates the Lense-Thirring 
precession of the nodes; therefore, the gravitomagnetic clock effect is absent 
in this case. Hence we assume $\a \neq \pi/2$ and consider $\tau_+$
from equation (\ref{2ini4}) such that $\varphi_0\rightarrow \varphi_0
+2\pi$ from $\tau=0$ to $\tau=\tau_+$; we find an implicit equation for 
$\tau_+$ that can be solved perturbatively. Thus let
\be
\tau_{\pm}\,=\,T\pm 2\pi\,\frac{a}{c}\,\lambda \cos\a\,;
\label{ansatz}
\ee
then, it is possible to show that 
\be
\lambda \,=\, \Gamma_0 - 2\Gamma_0^{-1} \tan^2\a\cos^2\h_0
\label{possible}
\ee
for $\a$ sufficiently different from $\pi/2$ such that the perturbative 
treatment remains valid. It is important to recognize that $\lambda$ 
could change sign and become negative for sufficiently large inclination
of the orbital plane with respect to the equatorial plane of the central
source. For instance, for $\a$ and $\h_0$ both near $\pi /4$, $\lambda$
could go through zero and change sign. The azimuthal period of the 
prograde orbit would then become shorter than the retrograde orbit.

The influence of the gravitomagnetic field of the source is in general
reflected in any orbital timing mechanism. For instance, let us recall that
$t$ is the proper time of static observers at spatial infinity. Thus if the
orbital motion about a rotating source is referred to static clocks at spatial
infinity, we find
\be
   t_{\pm} = T_0 \pm 2\pi\,\frac{a}{c}\,\Gamma_0\lambda'\,\cos\a  \, ,
\ee
where $\Gamma_0 T = T_0$ and $\lambda' = \lambda - 3\,\Phi_0\,\Gamma_0$ using 
equations (\ref{2ini1}) and (\ref{ansatz}). This consideration could possibly
be useful in the experimental determination of $J/M$ as well.

It follows from these considerations that in the approximation under 
consideration, where the rotation of the source is taken into account
only to first order, the azimuthal orbital period $\tau$ -- as compared to 
the proper Keplerian period $T$ -- would be larger (smaller)
due to orbital motion with $\lambda > 0$ 
in the same (opposite) sense as the rotation of the
source. For the case of a satellite around the Earth, this difference is
$\pm 2\pi a\lambda\cos\alpha/c$, where $2\pi a/c \approx 10^{-7}$ sec. 
A similar situation holds -- as noted in the previous paragraph -- even 
when the binary system is far away. It is interesting to contrast this
circumstance with the variation in the 
period of the Hulse--Taylor binary pulsar
PSR B1913+16. The binary period is monotonically decreasing by about 
$10^{-7}$ sec per orbit as a result of gravitational 
radiation damping, since the observations agree with the theoretically
estimated decay due to the emission of gravitational radiation by the 
binary system. To measure an effect of this size for the motion of a 
clock in orbit around the Earth is the observational challenge posed
by the gravitomagnetic clock effect (cf.\ section 6).

Though our approach has been macrophysical throughout, it is nevertheless
interesting to note that for gravitational orbits around a neutron, say,
$\tau_+ - \tau_-$ is equal to the Compton period of the neutron.

Finally, imagine standard clocks in orbit about a static astronomical source. 
The orbital period is defined by azimuthal closure as before. Suppose that
for two clocks starting at $\tau=0$ in opposite directions around the 
source we have $\tau_+ = \tau_-$. This degeneracy is removed once the source
rotates [cf. equation (\ref{ansatz})]. It is then expected that 
$\tau_+ - \tau_- = 4\pi a\lambda\cos\a/c$ would be the dominant relativistic
rotation-dependent term in general. 

\section{Holonomy}
As already mentioned in section \ref{sec2}, the gravitomagnetic clock 
effect can be considered as a gravitational analog of the 
Aharonov--Bohm effect. It is well known that the Aharonov--Bohm effect is 
closely connected to the concept of holonomy (see, for example, \cite{naka90},
section 10.5.3.): The phase shift observed in the Aharonov--Bohm
experiment can be obtained after the integration of $U(1)$--parallel transport
around a closed loop which surrounds a magnetic field. 
Mathematically, this phase shift is given by the holonomy attributed 
to a fibre bundle with base space $S^1$ (representing the closed loop),
fibre $U(1)$, and a $U(1)$--valued connection (the vector potential
that corresponds to the magnetic field).
An analogous construction can also be conceived in order to obtain the 
gravitomagnetic clock effect as a translational holonomy 
of a fibre bundle. In this case one
defines a fibre bundle with base space $S^1$ (representing the parameter 
space of the azimuthal coordinate $\varphi$), the real line $R$ as fibre, 
and an $R$--valued connection (describing the time shift of an 
orbiting clock due to gravitomagnetism). However, this construction is of 
limited use since the holonomy obtained is {\it not} directly related to 
gravitational holonomy.

Gravitational holonomy is a characteristic of exterior parallel transport on
spacetime. More precisely, it is attributed to an affine frame bundle which
takes spacetime as base manifold and is equipped with an affine connection
(cf. Appendix B). 
The gravitational holonomy itself splits into translational holonomy and 
rotational holonomy. In the limit of infinitesimally closed and contractible
loops these holonomies turn, respectively, into the torsion and curvature
of spacetime.  In general relativity, one is restricted to Riemannian 
parallel transport such that torsion vanishes and curvature becomes
Riemannian curvature. However, it should be noted that, in general, Riemannian 
parallel transport on a Riemannian manifold exhibits nonvanishing 
translational holonomy. This circumstance does not imply the
presence of nontrivial torsion, of course. Nevertheless, Petti \cite{pett86}
employed spacelike circular curves in the Kerr geometry in a significant
attempt to relate the angular momentum that occurs in the Kerr metric to
torsion via translational holonomy. Indeed, it was outlined in \cite{pett86} 
how to assemble a manifold with torsion from torsion-less Kerr
configurations in a manner similar to the assembly of a simplicial
Riemannian manifold from piecewise flat spaces. This interesting
undertaking deserves further investigation since -- as Petti \cite{pett86} has
observed -- a careful limiting process should be taken into account.

It follows from this that the gravitomagnetic clock effect does not give us 
any information on the local geometry of the Kerr metric that goes beyond
Riemannian geometry. This is not really surprising since an ideal standard
clock, if considered as a structureless, i.e.\ pointlike, test mass, does not
couple to torsion at all \cite{hehl85}; in fact, it measures 
proper time along timelike worldlines and is thus intimately tied to the 
metric concept of Riemannian geometry.

\section{Discussion}
NASA's Gravity Probe$\,$-A involved the sub-orbital flight of a rocket
carrying a
hydrogen maser clock; the main result of this experiment was an accurate test
of the gravitational redshift by Vessot {\it et al.}\ 
\cite{VesLev, vess80}. It is
expected that some of the future space missions will carry ultra-stable clocks,
and hence it is possible that the gravitomagnetic effect discussed in this
paper could be measurable in the future. For an orbit of Keplerian period 
$T_0$, equations (\ref{ansatz}) and (\ref{possible}) 
imply that the relative gravitomagnetic variation in orbital period is
\be
       \frac{\tau_+ - \tau_-}{T_0} \simeq \frac{2\,J}{M\,c^2}
       \left(\frac{G\,M}{\rho^3}\right)^{1/2}   \, ,
\label{relative}
\ee
or about $4\times 10^{-11}$ for a near-Earth orbit.

The principal aim of NASA's Gravity Probe$\,$-B will be the direct measurement
of the gravitomagnetic field of the Earth by comparing the precession of
gyroscopes with respect to telescopes on board a drag-free satellite in a polar
orbit about the Earth \cite{ever86}. After one orbital period, 
the gravitomagnetic
precession angle of the gyroscope per $2\pi$ radians is
\be
   \frac{\Omega_P}{\omega_0} \simeq \frac{2\,G\,J}{c^2\,\rho^3}
   \left(\frac{G\,M}{\rho^3}\right)^{-1/2}  \, ,
\ee
which is identical with equation (\ref{relative}). 
This equality indicates that the
theoretical possibility of measuring gravitomagnetism via clocks is essentially
equivalent to the GP-B.

The gravitomagnetic clock effect constitutes
a surprisingly large effect. We mentioned already in 
section \ref{sec2} that in the 
case of an astronomical body such as the Earth, the time difference 
$\tau_+ - \tau_- \simeq 4\pi J/Mc^2$ after one orbit amounts to
\be
\tau_+ - \tau_-\, \simeq \, 2\times 10^{-7}{\rm sec} \,.\label{oureffect}
\ee 
Time shifts due to the gravitomagnetic field of the Earth are 
widely believed to be of much lower order.
Indeed, suppose the time difference of a direct and retrograde moving
clock is taken at a {\it fixed time}, say, after one Kepler period
$T_0=2\pi /\omega_0$. Then it is straightforward to show that for
a circular equatorial orbit,
\be
{\tau'}_+ - {\tau'}_- \, = \, 12\pi\frac{GJ}{c^4r} + {\cal O}(c^{-6}) \,.
\ee
As an example, we set $r = 7000\, {\rm km}$ and obtain 
\be
{\tau'}_+ - {\tau'}_- \, \simeq \, 3\times 10^{-16} {\rm sec}\,.
\ee
This is nine orders of magnitude smaller than the time difference 
(\ref{oureffect}). We recall that the result (\ref{oureffect})
presupposes that the time difference of the two clocks is taken with respect
to a {\it fixed angle $\varphi$} (i.e., after each clock has covered an 
azimuthal interval of $2\pi$) and not with respect to a fixed 
time.

In principle, it is a trivial task to measure a time 
difference of $2\times 10^{-7}{\rm sec}$ with today's technology. 
However, an experimental 
verification of the gravitomagnetic clock effect does not only require 
the measurement of the time difference between 
two well-defined events up to an
accuracy of $2\times 10^{-7}$ sec. It is, in fact, the proper time along the 
direct and retrograde orbits that is used as a clock. 
Therefore, it is not sufficient to send two highly accurate and stable clocks 
into space and let them orbit in opposite directions. The orbits 
themselves have to be highly accurate and stable as well. We recall that 
in order to obtain the time difference (\ref{oureffect}) we have to 
subtract two periods, each of which represents 
the sum of a Kepler period and a much smaller relativistic correction. 
Under the assumption of {\it identical} orbits, the Kepler periods and their
gravitoelectric corrections cancel upon subtraction while the gravitomagnetic
contributions add up, yielding the actual clock effect under consideration
here. Disturbances 
of the orbits will in general change the Kepler periods  
of the orbiting clocks. It follows that in this case the Kepler periods will 
not exactly cancel but may exhibit a significant time difference.

In order to obtain some estimation of possible errors, 
we consider an actual experiment 
where atomic clocks are on board satellites in direct and retrograde 
orbits around the Earth. We may divide 
the error sources of such an experiment  
into two groups, namely
\begin{itemize}
\item[(i)] errors due to the tracking of the actual orbits, and
\item[(ii)] deviations from idealized orbits due to
\begin{itemize}
\item{} mass multipole moments of the Earth
\item{} radiation pressure
\item{} gravitational influence of the Moon, the Sun, and other planets
\item{} other systematic errors (e.g., atmospheric disturbances).
\end{itemize}
\end{itemize}

The tracking of the actual orbits requires the measurement of distances 
and angles.  The position of a satellite along an orbit can be determined to
a few centimeters using the Global Positioning System (GPS); therefore, the
temporal uncertainty that a near-Earth 
satellite has actually returned to the same
azimuthal position in space is about
$\delta\,\tau \approx \delta\,r/v \approx 10^{-6}\,{\rm sec}$. Here
$\delta\,r \approx 1\,{\rm cm}$ is the position 
uncertainty along track and $v$ is
the orbital speed of the satellite. The gravitomagnetic clock effect, however,
involves a definite temporal deviation of $\sim 10^{-7}\,{\rm sec}$. 
It turns out that one should be able to measure the orbital radius up to 
an accuracy of the order of $10^{-2}\,$cm and to determine angles up to 
an accuracy of $10^{-10}\,$rad in order to 
keep the errors due to the measurement smaller
than the clock effect after one orbit. These requirements are about one order
of magnitude higher than what can be achieved today. 
On the other hand, the clock effect 
is cumulative -- just like the precession angle of a GP-B gyroscope -- and 
hence many orbital periods can be used for a measurement of the 
gravitomagnetic effect; that is, the statistical tracking errors could be
overcome if one were able to perform many single measurements \cite{grate}.

The astrometric requirements for ensuring azimuthal closure should be
emphasized. The operational definition of the azimuthal angle $\varphi$ 
is ultimately based on the underlying astronomical coordinate system
employed \cite{kope90}. In fact, it may turn out to be necessary to monitor 
orbital motion so accurately as to be able to measure the Lense--Thirring 
effect and hence gravitomagnetism directly using only astrometric data
on the orbits of spaceborne clocks.

The systematic errors of the second group of error sources 
have a more serious influence 
on the gravitomagnetic clock effect. In order to calculate the influence 
of such perturbative accelerations on the Kepler period one has to 
focus on sections of orbits rather than on complete closed 
orbits. This is because under favorable conditions (e.g., an almost 
constant perturbative acceleration) different temporal deviations can cancel 
each other when summed over a closed orbit. 
It is possible to estimate that perturbative accelerations should be 
kept below $10^{-11}\,$g in order for the 
clock effect to become measurable.

Perturbative accelerations due to multipole moments of the Earth are of the 
order of $10^{-3}\,$g. The orbit of a satellite under the influence 
of the nonspherical and inhomogeneous form of the Earth may be likened 
to a bumpy road. However, we do know the gravitational field of the Earth 
up to an accuracy of $10^{-9}\,{\rm g} - 10^{-10}\,$g. 
NASA's already approved gravity mapping mission GRACE  
is expected to push this accuracy higher by about two orders of magnitude  
\cite{priv}. 
This would then make it possible, in principle, to correct 
for the influence of the multipole moments of the Earth 
on a gravitomagnetic clock experiment. 

The radiation pressure of the Sun causes perturbative accelerations of
the order of $10^{-8}\,$g. Using drag-free satellite techniques, this
disturbance can be reduced by two orders of magnitude to $10^{-10}\,$g.
To keep this error source under control one thus has to be able
to determine the solar radiation pressure accurately enough such that 
corrective calculations can be performed. Alternatively, one must wait 
for drag-free satellites that
can perform at least one order of magnitude better than current technology.

The gravitational fields of the Moon 
and the Sun cause relative accelerations  
between the Earth and the orbiting clocks. The amplitudes
of these accelerations are of the order of
$10^{-7}\,$g (Moon) and $10^{-8}\,$g (Sun). The influence of the other
planets of the solar system plays only a minor role. For Jupiter, e.g.,
we obtain an influence of the order of $10^{-12}\,$g. The positions of the
Moon and the Sun are known with much higher accuracy than is needed to 
determine their gravitational field at the level of $10^{-11}\,$g;
therefore, in principle, the influence of the Moon and the Sun can be 
properly taken into account \cite{grate2}. 

It should be remembered that we have provided analytic expressions for 
$\tau_\pm$ only for equatorial circular orbits and for spherical 
orbits with definite inclination. Our treatment must therefore be extended
to eccentric orbits. This will probably require numerical 
integration of the corresponding geodesic equations. One should keep in 
mind that even initially circular orbits will acquire eccentricity
over time due to the perturbative accelerations mentioned above.
It is expected from these considerations and the previous discussion 
that the measurement of this gravitomagnetic effect would be similarly 
complicated as in the GP-B.

Finally 
it should be remarked that the possibility of sending electromagnetic signals
around a rotating mass has been considered by a number of authors 
\cite{davi74}. The
time difference, according to the observer's clock, for a signal to follow a
closed path (by means of ``mirrors'', etc.) in direct and retrograde directions
around a rotating mass is proportional to the gravitomagnetic flux threading
the loop. Similar effects arise in the synchronization gap around a rotating
mass. One can show that all such effects are smaller than the clock effect
considered in this paper by a factor of the order of $\Phi = G\,M/c^2\,r\,$;
$\Phi$ is $< 10^{-9}$ for the exterior field of the Earth and $< 10^{-6}$ for
the exterior field of the Sun \cite{cohe93}. The clock effect is larger than
that for light by the inverse of a factor of the 
order $v^2/c^2\sim\Phi\,$; intuitively, the clock accumulates a much larger 
gravitomagnetic effect since its Keplerian motion
has a speed that is much smaller than the speed of light.

\section*{Acknowledgments}
Thanks are due to Ralph Metzler for his assistance with the preparation
of the figures. Helpful discussions with Adam Helfer, Sergei Kopeikin, and
Gerhard Sch\"afer are gratefully acknowledged. This work has been supported 
in part by the Alexander von Humboldt Foundation.

\appendix
\section*{Appendix A: Gravitational Larmor Theorem}
The local Larmor equivalence between magnetism and rotation has an exact analog
in the theory of gravitation. To elucidate this connection, let us consider the
motion of free test particles in the gravitational field given by the metric
form (1). The geodesic motion follows from $\delta\int -m c\,ds = 0\,$;
however, for purposes of comparison with the Newtonian equations of motion
\cite{newton}, we
consider $\delta\int L\,dt = 0$ with $L = -m c\,ds/dt\,$. That is
$$
       L = - m\,c^2\,\left[ 1 - 2\Phi - (1+2\Phi)\,\frac{v^2}{c^2} +
           \frac{4}{c^3}\,{\bf v}{\bf \cdot}{\bf A}_g\right]^{1/2}  \, .
                                                                \eqno({\rm A}1)
$$
It follows from this Lagrangian that the canonical momentum is given by
$$
   {\bf P} = m\,\Gamma\,(1 + 2\,\Phi)\,\frac{d{\bf x}}{dt} -
        \frac{2\,m}{c}\,\Gamma\,{\bf A}_g  \, ,             \eqno({\rm A}2)
$$
where $\Gamma = c\,dt/ds$ and $\Gamma \approx 1$ for nearly Keplerian motion
under consideration here. The analog of equation (A1) in electrodynamics is
$$
L= -mc^2\Bigl(1-\frac{v^2}{c^2}\Bigr)^{\frac{1}{2}}+\frac{q}{c}(-c\phi
+{\bf v}{\bf \cdot}{\bf A})\,,    \eqno({\rm A}3)
$$
where $\phi$ is the electric potential. A perturbative treatment of equation 
(A1) can bring it to the form of equation (A3) with $q\rightarrow -m$ except
for an extra factor of 2 in the term involving the gravitomagnetic vector 
potential; indeed, the origin of this extra factor is that off-diagonal 
terms appear in the spacetime interval with a factor of 2. The simplest way 
to deal with this situation is to assign a gravitoelectric charge of $-m$
and a gravitomagnetic charge of $-2m$ to the test particle. 
In this way, equation (A2) becomes consistent with the standard
relationship in electrodynamics between the canonical momentum ${\bf P}$ and
the kinetic momentum ${\bf p}$ of a particle of charge $q$ in an inertial
frame in Minkowski spacetime, ${\bf P} = {\bf p} + q\,{\bf A}/c\,$. It is
convenient to use units such that $G=1$ in what follows. In these units,
let us therefore assign gravitomagnetic charges of $q_B = - 2\,m$ to the test
particle and $Q_B = 2\,M$ to the central source. The source is rotating;
therefore, it has a gravitomagnetic dipole moment
${\boldsymbol \mu}_g = Q_B\,{\bf J}/2\,M\,c = {\bf J}/c$ 
and hence a gravitomagnetic
vector potential ${\bf A}_g = {\boldsymbol \mu}_g \times 
{\boldsymbol \rho}/\rho^3\,$. This is
consistent with equation (2); hence, the analogy established thus far between
magnetism and gravitomagnetism is exact in the approximation under
consideration here. That is, the gravitomagnetic dipole moment for a test
gyro of spin ${\bf S}$ is then $\widehat{\boldsymbol \mu}_g =
-{\bf S}/c$ and the rate of precession of this dipole moment in the 
gravitomagnetic field, $d{\bf S}/dt = \widehat{\boldsymbol \mu}_g 
\times {\bf B}_g$, is precisely given by equation (\ref{precess}).  
Moreover, the comparison of equation (A1) with its
electrodynamic analog (A3) leads to the interpretation that the factor of $2$ 
in the gravitomagnetic charge has to do with the fact that $ds^2$ is a 
quadratic form or, equivalently, that gravitation is a spin-2 field.

The correspondence between magnetism and gravitomagnetism can be simply
extended to the Larmor theorem. Thus the gravitomagnetic Larmor frequency is
given by
${\boldsymbol \Omega}_L = q\,{\bf B}_g/2\,m\,c = 
- {\bf B}_g/c = - {\boldsymbol \Omega}_P\,$.
The gravitational Larmor theorem obviously holds for the gravitomagnetic
precession of a gyroscope, since it means that the same rate of precession
would be obtained {\it locally} for a free ideal gyroscope in the absence of
gravitation but observed from a frame rotating with frequency
${\boldsymbol \Omega}_L = - {\boldsymbol \Omega}_P\,$. Further 
applications of these ideas are
contained in \cite{mash93}.

In view of the intrinsic significance of the gravitational Larmor theorem and
its basic relationship with Einstein's principle of equivalence, it is 
important to point out some of its underlying features. The theorem is
formulated with respect to preferred standard observers in the background
Cartesian coordinate system, i.e. the rest frame of the center of mass of the
rotating source. Furthermore, it has been assumed in our discussion thus 
far that gravitomagnetism arises from {\it mass currents}, as in the case
of a rotating planet or star, so that the gyromagnetic ratio is then unity.
In this case, $q_E \neq q_B$ and hence Lorentz invariance is broken. It is 
necessary to point out, however, that our treatment can be extended to the
gravitomagnetic field of a rotating black hole as well.

Consider first a test gyro in the field of an arbitrary rotating configuration.
It is reasonable to suppose that the gravitomagnetic field is independent 
of the specific nature of the source in the linear approximation under 
consideration here. Hence, let ${\boldsymbol \mu}_g =\kappa{\bf J}/c$
for the source and $\widehat{\boldsymbol \mu}_g = -\kappa{\bf S}/c$
for the test gyro, where $\kappa$ is a universal constant regardless of the
nature of the rotating systems. Then, $d{\bf S}/dt = 
\widehat{\boldsymbol \mu}_g \times {\bf B}_g$ implies that 
$\kappa^2=1$ and we choose $\kappa=1$ by convention and without any loss 
in generality. Next, let $\mu=\gamma QJ/2Mc$, where $\gamma$ is the 
gyromagnetic ratio and $\kappa=\gamma Q_B/2M=1$. In view of our previous 
arguments, cf.\ equations (A1)--(A3), $Q_B=2M$ and $\gamma=1$; however, in
the absence of such arguments the simplest possibility is $Q_E=Q_B=M$ and
hence $\gamma=2$. In fact, the charged Kerr system has $\mu = Qa$ and 
$\gamma=2$. In this way, the gravitational Larmor theorem can be extended
to the exterior field of a Kerr system. Our results suggest, but do not 
prove, that $\gamma=2$ for the Kerr-Newman system since gravity is a 
spin-2 field. That is, the factor of 2 in the gravitomagnetic charge
of a rotating mass that is connected with the spin-2 character of the 
linearized gravitational field simply goes over to the gyromagnetic ratio 
for the case of a charged Kerr configuration. This apparent independence 
of $\gamma$ from the detailed electromagnetic nature of the source for this 
case is supported by the circumstance that $\gamma=2$ extends to the 
{\it generalized Kerr--Newman spacetime} \cite{quev91}, which contains an 
infinite set of multipole moments, indicating the insensitivity of the 
gyromagnetic ratio to the detailed structure of the source. Further evidence 
for this viewpoint comes from the fact that a proper classical interior 
electromagnetic source for the exterior Kerr--Newman field should have 
$\gamma=1$ {\it in the absence of gravity}. In fact, the gyromagnetic ratio of
the Kerr--Newman system loses its meaning once one isolates the electromagnetic
source of the Kerr--Newman geometry by simply turning the gravitational 
interaction off; that is, one is then left with only a charge $Q$ and a 
static magnetic dipole moment $Qa$. 

\section*{Appendix B: Gravitational Holonomy}

In general relativity the {\it local} geometry of spacetime is determined 
by a metric $g$. Parallel transport in spacetime is then described by
means of a Riemannian connection one--form $\rG_\a{}^\b$ which is 
derived from the metric $g$. The prescription of parallel transport is the 
following: Consider a linear frame $e_\a$ at a point $x$ of spacetime. Parallel
transport of $e_\a(x)$ to an infinitesimally neighboring point $x' = x+dx$
is defined by the identification of the frame
$$
(e_\a + de_\a)(x') = e'_\a+de_\a(x'):= e'_\a + \rG_\a{}^\b(x') e'_\b
 \eqno({\rm B}1)
$$
with the parallel transported frame $e_\a(x)$.\footnote{This definiton 
does depend on the choice of $e'_\a=e_\a(x')$, of course. Another choice 
of $e'_\a$ would correspond to another gauge of $\rG_\a{}^\b$.}  
It is well known that this Riemannian parallel transport is both metric
preserving and torsion--free. The Riemannian curvature two--form 
$$
\rR_\a{}^\b := d\rG_\a{}^\b - \rG_\a{}^\g\wedge\rG_\g{}^\b
\eqno({\rm B}2)
$$
is a local measure of the nonintegrability of Riemannian parallel transport. 
That is, if we parallel transport a frame around an {\it infinitesimal} 
closed loop in spacetime the frame will undergo a ``pseudo-orthogonal'' 
linear transformation which is determined by $\rR_\a{}^\b$.
We can generalize this concept and consider parallel transport 
of a frame around a finite loop. This will again result in a 
``pseudo-orthogonal'' linear transformation of the frame and constitutes the
holonomy. More exactly, it is the holonomy of a linear frame bundle with
spacetime as base manifold and a Riemannian connection as connection.

More general non-Riemannian spacetime geometries are determined not only
by a metric $g$ but also by an independent affine connection which usually
can be taken as a {\it Cartan connection} ($\vt^\a, \G_\a{}^\b$)
\cite{cart86}. The 
Cartan connection determines affine parallel transport which generalizes
Riemannian parallel transport. While Riemannian parallel transport determines
the identification of neighboring linear frames by means of 
(pseudo-)ortho\-gonal transformations, the affine parallel transport 
determines 
the identification of neighboring affine frames by means of affine 
transformations \cite{koba63}. Note that an affine frame 
at a point $x$ of a spacetime $M$ is a pair 
$(e_\a, p)(x)$ which consists of a linear frame $e_\a(x)$ and a point $p(x)$. 
The affine frame is defined as an element of the affine tangent space $A_x M$.
We recover from the affine frame $(e_\a, p)(x)$ the linear frame $e_\a(x)$
if and only if $p\in A_x M$ is identified with $x$.   

The precise prescription of affine parallel transport is the following:
Consider a linear frame $e_\a$ at a point $x$ of spacetime and 
define $\G^{(T)\a}$ as $\G^{(T)\a}:= \vt^\a - \d^\a_i dx^i$. 
Affine parallel transport of $(e_\a, p)(x)$ to an infinitesimally 
neighboring point $x' = x+dx$ is defined by the identification of the affine 
frame
$$
(e_\a+de_\a)(x')={e'}_\a+de_\a(x') := {e'}_\a + \G_\a{}^{\b}(x')e'_\b 
\in A_{x'}M 
\eqno({\rm B}3) \label{cart1}
$$
$$
(p+dp)(x') ={p'}+dp(x') := 
{p'}+\G^{(T)\a}(x') \,{e'}_\a \in A_{x'} M
\eqno({\rm B}4) \label{cart2}
$$
with the parallel transported frame $(e_\a ,p)(x)$.
Integration of (B3) and (B4) around an infinitesimal 
closed loop yields the torsion and curvature two-forms $T^\a$ and 
$R_\a{}^\b$. The explicit defining formulas are
$$
{\rm torsion}\quad\qquad T^\alpha:={D}\vartheta^\alpha=
d\vartheta^\alpha+\Gamma
_\beta{}^\alpha\wedge\vartheta^\beta \,,\label{torsion}
\eqno({\rm B}5)
$$
$$
{\rm curvature}\qquad R_\alpha{}^\beta :=
d\Gamma_\alpha{}^\beta-\Gamma_\alpha{}^\gamma\wedge\Gamma_\gamma{}^\beta
\,.\label{curvature} \hspace*{1.16cm}
\eqno({\rm B}6)
$$
Torsion and curvature determine the resulting affine transformation of
an affine frame after parallel transport 
around an infinitesimal closed loop. The torsion
measures the translational part of this transformation while the curvature
measures the homogeneous part. 
Generalizing affine parallel transport to parallel transport around finite
loops yields the translational and rotational holonomy which we referred 
to as gravitational holonomy. It is the holonomy related to an affine 
frame bundle with spacetime as base manifold and a Cartan connection
as connection. Further discussion of gravitational holonomy can be found,
for example, in references \cite{scho, stell80}.

\end{document}